\documentclass[aps,prl,showpacs,twocolumn,preprintnumbers]{revtex4}
\usepackage{graphicx}
\usepackage{bm}
\usepackage{amsmath}

\def\be{\begin{equation}}       \def\ee{\end{equation}}
\def\bea{\begin{eqnarray}}      \def\eea{\end{eqnarray}}

\begin{document}

\title{Quantum Spin Hall Effect and Enhanced Magnetic Response by Spin-Orbit 
Coupling}
\author{Shuichi Murakami}
\email[Electronic address: ]{murakami@appi.t.u-tokyo.ac.jp}
\affiliation{Department of Applied Physics, University of Tokyo,
Hongo, Bunkyo-ku, Tokyo 113-8656, Japan}

\begin{abstract}
We show that the spin Hall conductivity in 
insulators is related with a magnetic susceptibility 
representing the strength of the spin-orbit coupling.
We use this relationship as a 
guiding principle to search real materials 
showing quantum spin Hall effect.
As a result, we theoretically predict that 
bismuth will show the quantum spin Hall effect, 
both by calculating the helical edge states, and by 
showing the non-triviality of the $Z_2$ topological number,
and propose possible experiments.\end{abstract}
\preprint{NSF-KITP-06-54}
\maketitle
Spin Hall effect (SHE) \cite{dyakonov1971b,murakami2003a, sinova2004}
has been attracting much attention recently,
partly due to potential use for semiconductor spintronics.
Its remarkable feature is to induce
a spin current without breaking 
time-reversal symmetry.
One of the interesting proposals is 
the quantum spin 
Hall (QSH) phase \cite{kane2005,bernevig2005a,qi2005,%
onoda2005a,wu2005,xu2005}, which is a 2D insulator with helical
edge states. The edge states form Kramers pairs, with
spin currents flowing oppositely for opposite directions of spins. 
The QSH phase can be regarded as a novel phase constrained by 
the 
$Z_2$ topological number $I$ \cite{kane2005}.
It is equal to a number of 
Kramers pairs of helical edge states modulo two.
The QSH
phase has $I=\mathrm{odd}$, while 
the spin-Hall-insulator (SHI) phase \cite{murakami2004c}, 
topologically equivalent to a simple insulator,
has $I=\mathrm{even}$.
It is surprising that insulators without ordering, usually considered as 
featureless and uninteresting, can have a
nontrivial topological QSH phase. Its nontriviality reveals itself e.g. in 
a critical exponent \cite{onoda2006}.
For its interest akin to the quantum Hall systems, 
an experimental observation of 
the QSH phase is called for.

To search for
candidates for the QSH phase among a vast number 
of nonmagnetic insulators, we need a guiding principle.
In this paper, we 
propose that the magnetic susceptibility would be a good 
measure, and 
we pick up bismuth as a candidate
due to its strong diamagnetism. There are also
other supporting clues:
large spin splitting in the surface states of the 3D system,
similar to edge states for the 2D systems, the crystal 
structure in the (111) plane 
similar the Kane-Mele model for the QSH phase \cite{kane2005}. 
Using a 2D tight-binding model,
we show that this system has only one pair of
edge modes. Its $Z_2$ topological number 
\cite{kane2005} is shown to be odd, i.e. nontrivial, which supports 
stability of the edge state.
These aspects make bismuth a promising candidate for the QSH phase.

To relate magnetic susceptibility with spin Hall conductivity
(SHC),
we derive here a spin-Hall analog of the St\v{r}eda formula. 
The St\v{r}eda formula \cite{smrcka1977,streda1982} 
tells us that in insulators
 the Hall conductivity $\sigma_{xy}$ is expressed as
$\sigma_{xy}=\frac{e}{\Omega}\frac{d N}{d B}|_{\mu}$, where 
$N$ is the number of states below the chemical potential $\mu$, 
and $\Omega$ is the area of the system.
A direct generalization to the SHC is to replace
$N$ by the spin $s_z$, i.e. 
$\sigma_{s}=\frac{1}{\Omega}\frac{d s_{z}}{d B}|_{\mu}$. This
is physically reasonable from the following argument.
Suppose we apply a magnetic field in some region, linear in 
time. Then a change of the total 
spin inside the region is proportional to 
$\frac{d s_{z}}{d B}|_{\mu}$. According 
to the Maxwell equation, the increasing magnetic field induces a 
circulating electric field. Therefore, by interpreting the 
spin change as due to a spin Hall current by the electric field, 
the SHC is $\sigma_{s}=
\frac{1}{\Omega}\frac{d s_{z}}{d B}|_{\mu}$. 
We can justify this argument by explicit calculations.

A key step is to use a ``conserved'' spin current \cite{shi2006}.
In most papers on the SHC, the spin current is conventionally defined
as 
$\vec{J}_s=\frac{1}{2}
\{\vec{v},s_z\}$. 
However, 
in the presence of the spin-orbit coupling, 
the spin is no longer conserved: ${\cal T}\equiv \dot{s}_{z}\neq 0$.
Namely, the rhs of the equation of continuity,
$\partial_t s_z+\vec{\nabla}\cdot \vec{J}_s=
{\cal T}$,
is nonzero, and
the ``conventional'' spin 
current $\vec{J}_s$ is not directly related with spin accumulation.
Instead, Shi {\it et al.} \cite{shi2006} defined a conserved spin current
$\vec{{\cal J}}_{s}$ as follows. If the system satisfies $\int dV
{\cal T}=0$, as it does in a uniform electric field, one can 
write ${\cal T}$ as ${\cal T}=-\vec{\nabla}\cdot 
\vec{P}_{\tau}$. Thus 
a conserved spin current defined as 
$\vec{\cal J}_{s}\equiv 
\vec{J}_s+\vec{P}_{\tau}$ satisfies
$\partial_t s_z+\vec{\nabla}\cdot \vec{\cal J}_s=0$,
and is calculated for several models \cite{shi2006,sugimoto2006}.

To calculate the SHC for the conserved spin current, 
we consider an electric field
with wavenumber $q$, and take the limit 
$q\rightarrow 0$ \cite{shi2006}. When we calculate a spin current 
flowing to the $x$-direction in response to an electric field
to the $y$-direction, we take 
the vector potential $\vec{A}=A_y e^{iqx}\hat{y}$, and  
the response is calculated 
as
\begin{eqnarray}
&&\sigma_{s}^{\cal J}=-\frac{e}{\Omega}\lim_{q\rightarrow 0}i\partial_{q}
\sum_{n\neq m}\frac{f(\varepsilon_{m})-f(\varepsilon_{n})}{
(\varepsilon_{n}-\varepsilon_{m})(\varepsilon_{n}-\varepsilon_{m}+i\eta)}\nonumber \\
&&\ \ 
\langle n|[H,s_{z}e^{iqx}]|m\rangle\langle m
|\frac{1}{2}\{v_y, e^{-iqx}\}|n\rangle. 
\end{eqnarray}
By rewriting as
$[H,s_z e^{iqx}]=\frac{1}{2}\{s_z,\ [H,e^{iqx}]\}+
\frac{1}{2}\{[H,s_z],\ e^{iqx}\}$, the first and the second term correspond
to the conventional 
and spin-torque terms ($\sigma_{\mu\nu}^{s0}$ and
$\sigma_{\mu\nu}^{\tau}$
in Eq.~(10) of Ref.~\cite{shi2006}), respectively.
By a calculation similar to \cite{yang2005},
we
get $\sigma_{s}^{\cal J}=\sigma_{s}^{{\cal J}(I)}+
\sigma_{s}^{{\cal J}(II)}$ with
\begin{eqnarray}
&&\sigma_{s}^{{\cal J}(I)}=
\frac{ie}{8\pi\Omega}\int d\varepsilon 
\frac{df}{d\varepsilon}
\mathrm{tr}\left(\left(
[H,s_{z}]G_{+}\{v_y, x\} \right. \right.
\nonumber \\
&&\ \ \ -\{v_y, x\}G_{-}
[H,s_{z}]
-2[H,s_{z}x]G_{+}v_y
\nonumber\\
&&\ \ \ \left. \left.+2v_{y}G_{-}
[H,s_{z}x]+[x,[s_{z},v_{y}]]
\right)(G_{+}-G_{-})\right),\nonumber \\
&&\sigma_{s}^{{\cal J}(II)}=
\frac{ie}{\Omega}\int \frac{d\varepsilon}{4\pi}
f(\varepsilon)\mathrm{tr}\left(s_{z}G_{+}L_z G_{+}
-G_{-}L_{z}G_{-}\right),
\end{eqnarray}
where $G_{\pm}=(\varepsilon-H\pm i\eta)^{-1}$ and 
$L_{z}= xv_{y}-yv_{x}$ is an orbital angular momentum.
The term $\sigma_s^{{\cal J}(I)}$ is proportional to 
$\frac{df}{d\varepsilon}(G_{+}-G_{-})$. 
By noting $G_{+}-G_{-}=-2\pi i\delta (\varepsilon-H)$, 
only the states at the Fermi energy contribute to 
$\sigma_s^{{\cal J}(I)}$. 
In insulators $\sigma_s^{{\cal J}(I)}$ vanishes identically.
On the other hand, the second term  $\sigma_s^{{\cal J}(II)}$ 
is expressed as
\begin{equation}
\sigma_{s}^{{\cal J}(II)}=
\int \frac{d\varepsilon}{\Omega}
f(\varepsilon)\mathrm{tr}\left(s_{z}\frac{d\delta(\varepsilon-H)}{dB_{
\mathrm{orb}}}
\right)=\frac{1}{\Omega}\frac{ds_{z}}{dB_{\mathrm{orb.}}}.
\label{streda1}\end{equation}
Equation (\ref{streda1}) agrees with the
above-mentioned physically expected form.
This result (\ref{streda1}) can be also written as
\begin{equation}
\sigma_{s}^{{\cal J}(II)}
=\frac{-\hbar}{g\mu_B}
\frac{1}{\Omega}\frac{dM_{\mathrm{orb.}}}{dB_{\mathrm{Zeeman}}}
=\frac{1}{\Omega g}\frac{dL_{z}}{dB_{\mathrm{Zeeman}}}.
\label{streda2}
\end{equation}
where $g$ 
is the electron-spin 
$g$-factor, $M_{\mathrm{orb.}}$ is an orbital magnetization,
and $\mu_{B}$ is the Bohr magneton.
These formulae are a spin analog of the St\v{r}eda formula 
\cite{streda1982}. 
We note that a St\v{r}eda formula for the
SHC
with the conventional
spin current $\vec{J}_{s}$ \cite{yang2005}
has extra terms involving $\dot{s}_{z}$, in addition to 
(\ref{streda1}). These terms arise from spin nonconservation.
Hence it is natural that they do not appear
in (\ref{streda1}), as we used the conserved spin current \cite{note}.
We remark that 
the definition of the spin current is still controversial.
Since the spin is not conserved, there is no unique definition of the 
spin current. Because there is no established way of directly measuring 
the spin current, one way is to consider instead a measurable 
quantities such as spin accumulation at edges.
The spin accumulation depends crucially on boundary conditions, and 
the conserved spin current may correspond to
smooth boundaries \cite{shi2006}. This point requires further investigation.

\begin{figure}
\includegraphics[scale=0.4]{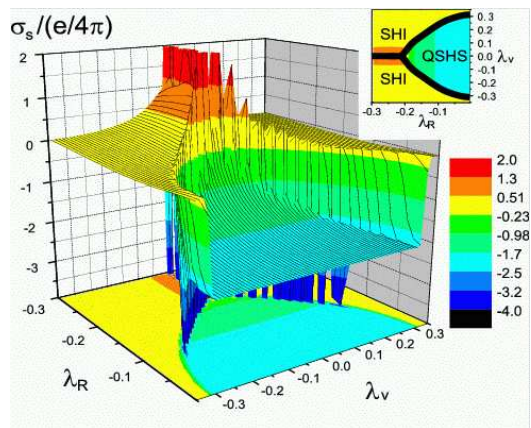}
\caption{Spin Hall conductivity $\sigma_s$ 
 for the Kane-Mele model on the honeycomb 
lattice for various values of $\lambda_R$ and $\lambda_{v}$
in the unit of $t$. $\lambda_{\mathrm{SO}}$ is fixed as $0.06t$. 
As it is symmetric with respect to $\lambda_R\rightarrow -\lambda_R$, 
we show only the result for negative $\lambda_R$. 
The phase diagram is shown in the inset.
Except for the vicinity of the phase boundary, $\sigma_s\sim 0$
in the SHI phase, and $\sigma_s\sim -e/(2\pi)$ in the QSH phase.}
\label{fig:vr}\end{figure}

We calculate $\sigma_s$ for 
the model on the honeycomb lattice proposed by Kane and Mele
\cite{kane2005}. 
This model shows the SHI and the QSH phases, depending on 
parameters
$\lambda_R$, $\lambda_v$ and $\lambda_{SO}$. 
We numerically evaluate the SHC by 
Eq.~(\ref{streda2}) using 
the formula of orbital magnetization
\cite{xiao2005,thonhauser2005},
\begin{equation}
M_{\mathrm{orb.}}=\frac{ei}{2\hbar}\sum'_{n}
\int
 \frac{d^d k}{(2\pi)^{d}}
\left\langle
\frac{\partial u_{n\mathbf{k}}}{\partial \mathbf{k}}\right|\times
(2\mu-\varepsilon_{n\mathbf{k}}
-H)\left|\frac{\partial u_{n\mathbf{k}}}{\partial \mathbf{k}}\right
\rangle,
\label{Morb}
\end{equation}
where $\sum'_{n}$ is a sum over occupied bands.
The result is shown in Fig.~\ref{fig:vr}.
 The calculated SHC is 
$\sigma_s\sim 0$ in 
the SHI phase and $\sigma_s\sim\frac{-e}{(2\pi)}$ in the QSH 
phase, except for the vicinity of the phase boundary. 
The SHC in the QSH phase, $\sigma_s\sim -e/(2\pi)$, is interpreted as
a fundamental unit $e/(4\pi)$ 
times two, the number of the edge states. The quantization 
is exact when $s_z$ is a good quantum number, i.e. $\lambda_R=0$,
where
the system is a superposition of two quantum Hall systems
with $\sigma_{xy}=\pm e^{2}/h$.
Remarkably, even when $s_z$ is no longer conserved,
the SHC remains almost quantized. 
Deviation 
from the quantized value is more prominent near the phase 
boundaries, which is attributed to smallness of the  band gap. 

Therefore,
materials with large susceptibility would be a good 
candidate for the QSH phase. 
Namely, if the susceptibility is large, 
$\sigma_s$ should be large. Figure~\ref{fig:vr}
then suggests that the system should be either in the QSH phase or 
near the phase boundary to the QSH phase.
From this reason, we pick up some semimetals and related materials 
with large diamagnetic susceptibility, among which are bismuth 
and graphite.

Bi crystal has a rhombohedral structure, with trigonal symmetry 
around the (111) axis.
Bi is a semimetal with a small hole pocket at the $T$ point, and three
electron pockets at the $L$ points.
Its strong diamagnetism has been studied 
experimentally and theoretically. 
It is theoretically attributed to massive Dirac fermions at the $L$ 
and $T$ points \cite{Bi-diamag}. These Dirac fermions gives a 
diamagnetic susceptibility which is 
enhanced as  logarithm of the small energy gap \cite{Bi-diamag}.
Even when the Fermi energy is in the gap, this picture survives,
and the 
susceptibility becomes even larger.
Such Dirac fermions contribute to anomalous Hall effect and the SHE.
Hence, it is no wonder such enhanced diamagnetic susceptibility implies 
an enhanced charge/spin Hall conductivity.

\begin{figure}
\includegraphics[scale=0.2]{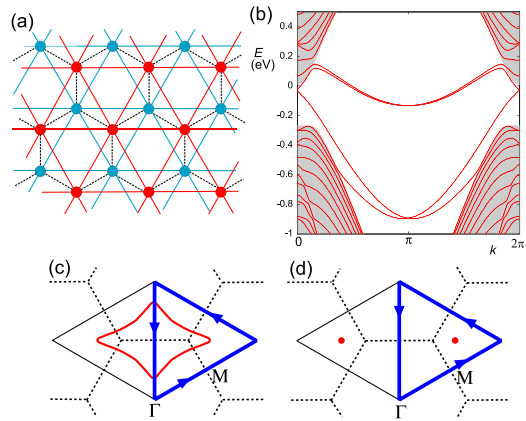}
\caption{Calculation on the bilayer tight-binding model of 
bismuth in the (111) plane. (a) the crystal structure of 
the (111)-bilayer bismuth. The upper and lower layers are 
denoted by red and blue, respectively. The solid and broken lines
represent intralayer and interlayer hoppings, respectively.  
(b) the calculated band structure  for the strip geometry 
with 20 sites wide.
The gray region is the bulk bands, while the red curves are the 
states calculated for the strip. 
All states are doubly degenerate.
Zeros of the Pfaffian $\mathrm{Pf} (\bm{k})$ in the Brillouin zone
are shown in red 
for (c) the 
inversion-symmetric
($v=0$) and for (d) the inversion-asymmetric ($v=0.2\mathrm{eV}$)
 cases.}
\label{Bi-bilayer}\end{figure}

Because the QSH phase is in 2D,
we have to make Bi two-dimensional,
such as thin films and quantum wells.
Such confinement
discretizes the perpendicular momentum, and tends to open the gap.
It was theoretically proposed that by making the Bi film thinner, 
it turns from 
semimetal to semiconductor \cite{lutskii,sandomirskii}.
Experiments show that the gap may open in thinner samples, 
whereas the gap is
obscured by carrier unbalance between holes and electrons
\cite{thin-film}.  
Remarkably, the lattice structure of Bi 
in the (111) plane resembles the Kane-Mele 
model \cite{kane2005}.
The 
crystal can be viewed as a stacking of bilayers along the [111] direction.
The inter-bilayer 
coupling is much smaller than the intra-bilayer
one, and
the LEED analysis showed that the (111) surface 
of Bi is terminated with an intact bilayer \cite{Moenig2005}.
Crystal structure of a single bilayer (Fig.~\ref{Bi-bilayer}(a))
consists of two triangular sublattices located in different layers. 
This honeycomb-like lattice structure is a key 
for a nontrivial $Z_2$ topological number; a crystal structure with 
high symmetry (e.g. square lattice) favors a trivial $Z_2$ topological 
number.

We demonstrate that 2D single-bilayer 
bismuth has a pair of helical 
edge states carrying spin currents with opposite spins.
Furthermore, we show that the $Z_2$ topological number is odd.
For these purposes, we 
use the 3D tight-binding model \cite{liu1995} which well
reproduces the band structure, and 
truncate the model by retaining only the hoppings inside the bilayer.
The resulting 16-band model is regarded as a multi-orbital
version of the Kane-Mele model.
We first calculate the band structure for a single bilayer 
(Fig.~\ref{Bi-bilayer}(a)) for a strip geometry. 
The result is shown 
in Fig.~\ref{Bi-bilayer}(b), where projected bulk bands are shown 
in gray. 
The figure shows four edge states connecting between 
the bulk conduction and valence bands. They 
correspond to one Kramers 
pair of edge states, 
suggesting nontrivial (odd) $Z_2$ topological number.
The spin Chern number \cite{spinCh} is calculated as $C_{sc}=-2$,
which is consistent with the existence of one pair of edge states.
We also calculate the spin Hall conductivity from 
Eq.~(\ref{streda2}) and we get $\sigma_s\sim 
-0.74\cdot\frac{e}{4\pi}$.
This value is reduced from $-2\cdot\frac{e}{4\pi}$ due to 
non-conservation of spin.

To confirm that the 2D bismuth is in 
the QSH phase, we also 
calculate the Pfaffian $\mathrm{Pf} (\bm{k})$
to calculate the $Z_2$ topological number
\cite{kane2005}.
The bilayer system is inversion-symmetric, allowing $\mathrm{Pf} (\bm{k})$
to be chosen real. The 
result is Fig.~\ref{Bi-bilayer}(c), where $\mathrm{Pf} (\bm{k})$ changes sign 
at the red curve, corresponding to Fig.~2 (a) in \cite{kane2005};
it implies odd $Z_2$ number.
To further clarify the phase winding of $\mathrm{Pf} (\bm{k})$, we break the 
inversion symmetry by adding small on-site energies $\pm v$, for the atoms 
on the upper and the lower layers, respectively. This may correspond to a
heterostructure or a single-bilayer thin film on a substrate.
The result is shown in Fig.~\ref{Bi-bilayer}(d) for $v=0.2 \mathrm{eV}$.
There is only one vortex for
the phase of $\mathrm{Pf} (\bm{k})$ in the half BZ, which ensures the odd 
$Z_2$ number, corresponding to Fig.~2(b) in \cite{kane2005}.
We note that 
the zeros of the Pfaffian do not follow
the threefold rotational symmetry.
It is because the Pfaffian is not covariant with respect to 
unitary transformation of the Hamiltonian, and 
depends on a choice of the unit cell. 
Because this QSH phase is protected by topology,
it cannot be broken unless 
the valence and conduction bands touch at the same wavenumber
and the direct gap closes.
Therefore, even though 
the 2D tight-binding model might not reproduce
quantitatively the real band 
structure, the nontriviality of the $Z_2$ index 
is more robust.
This nontrivial $Z_2$ number guarantees stability of the 
helical edge states \cite{wu2005,xu2005}.
Only when the $Z_2$ topological
number is odd, 
the edge states are stable against single-particle backscattering and 
(reasonably weak) two-particle backscattering, 
whereas for even topological number, the 
edge state will be gapped in general  \cite{wu2005,xu2005}. 
For example, in 
a bilayer antimony, we found that 
there are two pairs of edge modes,
and the $Z_2$ number is even. 
Thus the edge modes in 2D antimony 
will be fragile against opening a gap. 

We now discuss a multilayer Bi thin film.
Bismuth is suitable for pursuit of quantum size effects. 
A long mean-free path (
$l \sim \mu\mathrm{m}$-$\mathrm{mm}$) \cite{milimeter}, 
large electron mobility up to $10^6 \mathrm{cm}^{2}/\mathrm{Vs}$ at $5\mathrm{K}$ \cite{cho2001}, and a small effective mass of electrons make
the semimetal bismuth a good material to see quantum-size effects 
even at room temperature \cite{rogacheva2003}.
Furthermore, 
Bi thin films can be synthesized with good quality. For example, 
a  $10$ $\mu\mathrm{m}$-thick 
film shows magnetoresistance with a factor of few thousands
at $5\mathrm{K}$ and a factor of 2-3 at room temperature 
\cite{yang1999,cho2001}.

By stacking $N$ bilayers, the direct gap never closes, 
and each edge mode is topologically protected;
the number of pairs of edge modes becomes
$N$.
When the film becomes thicker, the helical edge states are expected to
evolve to 
2D surface states on a 3D bulk Bi, which 
has been studied by
the angle-resolved photoemission spectroscopy 
\cite{Ast2003a, Kim2005,Koroteev2004}.
The surface states with large spin splitting is observed 
\cite{Koroteev2004}, which manifests spin currents carried by such
surface states.
If the film thickness $D$ is less than the mean free path $l$,
e.g. $D\sim$ 10 $\mu\mathrm{m}$ \cite{yang1999,cho2001},
each of these edge modes has a
quantized motion in the perpendicular direction.
Hence, although the backscattering is relevant for even $N$, 
its effect on the edge states is almost negligible, and the 
system becomes gapless. By lowering the temperature or by 
increasing the disorder, the system will eventually become the SHI or 
the QSH phases depending on whether $N$ is even or odd.

To observe the QSH phase experimentally, one way is to measure 
the spin current by an 
applied electric field. 
We note that the QSH phase does not show a 
quantization \cite{kane2005}, unlike the 
 quantum Hall effect (QHE).
On the other hand, surprisingly, the critical 
exponent $\nu$, governing the localization length is found to be 
different between
the QSH and the SHI phases \cite{onoda2006}. 
Thus, the QSH phase can be established via measurement of 
$\nu$. 
In the QHE, $\nu$
is determined experimentally by changing the magnetic field 
across the plateau transition \cite{wei1988,koch1991a,huckestein1995},
because the change of the magnetic field controls the Fermi 
energy across the extended state. In the QSH phase, 
for example by a change of a gate voltage,
one may be able to 
control the Fermi energy to the extended or localized states.
To determine $\nu$, one has to see 
the range of the gate voltage $\Delta V_g$ showing nonzero $\sigma_{xx}$ 
by varying
the sample size \cite{koch1991a} or 
the temperature \cite{wei1988}. From their critical exponents, 
$\nu$ is calculated.

Another way to establish the QSH phase is to 
observe the edge states 
by  scanning tunneling microscopy/spectroscopy,
as has been used for graphite 
\cite{Niimi,Kobayashi}.
There is one important difference between the edge states 
of the graphite and the QSH phase.
In graphene the existence of edge states crucially depends on the edge shape;
the zigzag edge has edge states while the armchair edge does not. 
For a rough edge with
portions  
of zigzag and armchair edges, 
only at the zigzag edges can the signal of edge 
states be seen \cite{Kobayashi}. 
In contrast, the edge states in the QSH phase carry
helical spin 
currents, ad
circulate along the whole edge around, 
irrespective of the details (e.g. the shape) of the edge.

Besides bismuth, graphite is another material with 
anomalously large diamagnetic susceptibility \cite{McClure1956}. 
The spin-orbit coupling of graphite is
small, and the diamagnetism is mostly carried by 
orbital motion. In the 
same token as the SHE, 
the orbital-angular-momentum (OAM) Hall effect can be studied 
\cite{zhangyang2005}. 
The resulting OAM Hall conductivity is the 
susceptibility for the orbital:
$\sigma_{\mathrm{OAM}}^{{\cal J}(II)}
=\frac{1}{\Omega}\frac{dL_{z}}{dB_{\mathrm{orb.}}}$.
Since this involves only the orbital, the spin-orbit coupling 
is not required for it to be nonzero.  
Because the orbital susceptibility is largely enhanced in the graphite
due to massless Dirac fermions,
graphite will show large OAM current, 

In conclusion, we show that
the SHC is directly related with a ``spin-orbit'' susceptibility 
which is a response of the orbital magnetization by the 
Zeeman field.
We then propose that the magnetic susceptibility can be a good measure for 
search of quantum spin Hall systems. We theoretically 
predict that 2D bismuth will 
show the quantum spin Hall effect, because the number of 
pairs of 
helical edge states is odd and the $Z_2$ topological number is nontrivial. 
The sufrace states with large spin splitting in bulk bismuth,
might be closely related with these edge modes.

\begin{acknowledgments}
We thank B.~A.~Bernevig, M.-C.~Chang, M.~Onoda, S.~Onoda, 
N.~Nagaosa, and S.-C.~Zhang for helpful discussions. 
This research was supported in part 
by Grant-in-Aid from the Ministry of Education,
Culture, Sports, Science and Technology of Japan and 
by the National Science Foundation under Grant No. PHY99-07949.

\end{acknowledgments}


\begin{thebibliography}{99}
\bibitem{note}
In \cite{yang2005} it is commented, without giving details,
that even with the conserved spin 
current there appear extra terms (nonconserved part), which 
contradicts the present paper. 
One can easily check the calculations in the present paper,
and the reason for this difference is not clear.
\end{thebibliography}
\end{document}